\documentclass[twocolumn, trackchanges, tighten, times]{aastex62}
\shorttitle{Planet torques induced by scattered pebble-flow}
\shortauthors{Ben\'itez-Llambay, P. \& Pessah, M. E.}

\begin{document}

\title{\sc{Torques Induced by  Scattered Pebble-Flow in Protoplanetary Disks}}

\author{Pablo Ben\'itez-Llambay}
\email{pbllambay@nbi.ku.dk}

\author{Martin E. Pessah} \affil{Niels Bohr International Academy,
  Niels Bohr Institute, Blegdamsvej 17, DK-2100 Copenhagen \O{},
  Denmark}

\begin{abstract}
Fast inward migration of planetary cores is a common problem in the
current planet formation paradigm. Even though dust is ubiquitous in
protoplanetary disks, its dynamical role in the migration history of
planetary embryos has not been assessed. In this Letter, we show that
the scattered pebble-flow induced by a low-mass planetary embryo leads
to an asymmetric dust-density distribution that is able to exert a net
torque.
By analyzing a large suite of multifluid hydrodynamical simulations
addressing the interaction between the disk and a low-mass planet on a
fixed circular orbit, and neglecting dust feedback onto the gas, we
identify two different regimes, gas- and gravity-dominated, where the
scattered pebble-flow results in almost all cases in positive
torques. We collect our measurements in a first torque map for dusty
disks, which will enable the incorporation of the effect of dust
dynamics on migration into population synthesis models.
Depending on the dust drift speed, the dust-to-gas mass
ratio/distribution and the embryo mass, the dust-induced torque has
the potential to halt inward migration or even induce fast outward
migration of planetary cores.
We thus anticipate that dust-driven migration could play a dominant
role during the formation history of planets.
Because dust torques scale with disk metallicity, we propose that
dust-driven outward migration may enhance the occurrence of distant
giant planets in higher-metallicity systems.
\end{abstract}

	
\section{Introduction}

\begin{figure*}
  \includegraphics[width=\textwidth]{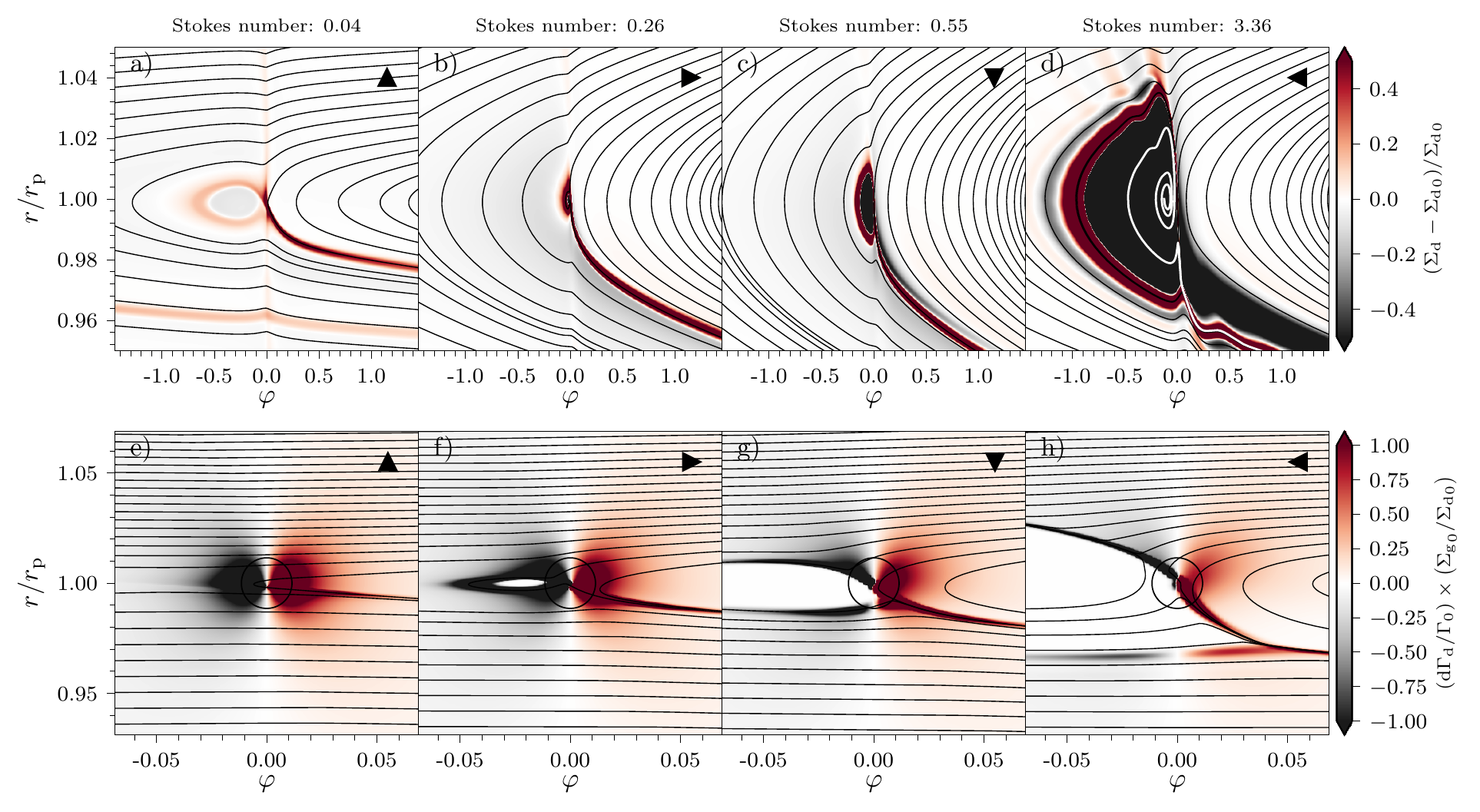}
  \caption{Top: relative dust-density perturbations and dust
    streamlines (black lines) for four different Stokes numbers,
    corresponding to the black triangles in Fig.\,\ref{fig:fig2}.
    Panel (d) includes a streamline starting inside the dust hole
    (white line) to illustrate the unstable behavior that leads to its
    formation. Bottom: associated torque density distribution,
    $d\Gamma_{\rm d}$ normalized by $\Gamma_0 = (m_{\rm
      p}/M_\star)^2{\Sigma_\mathrm{g}}_0 r_{\rm p}^4 \Omega_{\rm
      p}^2/h^{2}$. The black circle corresponds to the Hill radius of
    the planet. Note the different scales used in each set of panels.
    \label{fig:fig1}}
\end{figure*}

Planetary embryos are born and grow in a protoplanetary disk composed
of gas and a small fraction of dust, pebbles and rocky fragments. The
gaseous disk exerts a gravitational torque onto the nascent planets
that can modify dramatically their orbits \citep[e.g.,][]{Kley2012,
  Baruteau2013, Baruteau2014}.
Gradients in the density and temperature in the gas disk alter the
torques exerted on a planet, but have a mild impact on the so-called
differential Lindblad torque \citep{Ward1997}. An additional torque is
exerted by the horseshoe orbits around the protoplanet. The torque
associated with this so-called horseshoe drag \citep{Ward1991} can
become positive and larger than the differential Lindblad torque if
the gradients of vortensity or entropy are strong enough
\citep[e.g.,][]{Baruteau2008a,Masset2009,Paardekooper2010}. Typically,
the net torque resulting from these contributions is such that inward
migration is expected for low-mass embryos \citep{Tanaka2002a}.

While it is accepted that migration can be neglected for masses
smaller than the mass of Mars ($\sim 0.1 ~M_\earth$)
\citep[][]{Morbidelli2016a}, it becomes difficult to overcome inward
migration for embryos larger than $\sim 1 M_\earth$ on timescales of a
few Myr -- the lifetime of protoplanetary disks
\citep[e.g.][]{Mamajek2009}. In addition, planet population synthesis
studies usually require a reduced migration efficiency in order to
produce results consistent with observations
\citep[e.g.][]{Cossou2014}.
Several mechanisms able to reduce the migration rate have been
proposed during the last decade. Some of these are related to torques
in non-isothermal disks
\citep[e.g.][]{Paardekooper2006,Masset2010,Paardekooper2010,
  Paardekooper2011,Jimenez2017}, torques in magnetized disks
\citep[e.g.][]{Baruteau2011, Guilet2013, Uribe2015}, thermal torques
\citep[e.g.][]{Lega2014, Benitez-Llambay2015a, Masset2017,
  Masset2017b} and torques arising when considering the detailed
three-dimensional flow close to the planet \citep[e.g.][]{Fung2015a,
  Fung2017}. All of these effects rely on modifications to the
temperature and density of the gas in the vicinity of the planet.

Whereas the effects of gaseous torques on planetary migration have
been the focus of numerous studies, the possibility that
dust\footnote{Here, \textit{dust} refers to particles that are not
  directly subject to a dust pressure force but indirectly to the gas
  pressure by mean of a gas headwind.} could play a significant role
has been barely considered\footnote{\cite{Fouchet2010} seems to be the
  only work that has focused on this question. However, they did not
  find torques due to the dust component, perhaps due to the the large
  planetary mass considered.}.
This is perhaps not surprising as it is widely accepted that the
dust-to-gas ratio is only about $\sim 1$\% in mass, which makes it
hard to believe that dust can be dynamically important for migration.
In this Letter we show that dust-induced torques can indeed be
important because the dust-density distribution around the planet can
become significantly asymmetric due to the interplay between the drag
force and the scattering by the planet \citep[e.g., Figs. 2--5
  in][]{Morbidelli2012}.
Even though hints of these asymmetries in the dust distribution
surrounding the planet have already been observed
\citep[e.g.,][]{Dipierro2017}, the resulting dust-induced torque has
not been recognized and assessed in a systematic way.

We present systematic measurements of the torques produced by the dust
component of the disk and show that these are able to alter the
migration history of a protoplanet.
We show in the upper panel of Fig.\,\ref{fig:fig1} how these
asymmetries manifest themselves in the flow surrounding the
planet. For a wide range of conditions, the associated asymmetric
torque density distribution, shown in the lower panel of
Fig.\,\ref{fig:fig1}, leads naturally to a positive dust-induced
torque component.
By performing a suite of 200 numerical simulations, we find that
asymmetries in the scattered dust flow are able to change the
magnitude and even the sign of the net torque acting on a planet, and
that the value of the dust-induced torque depends on the particles'
size and mass of the planet. In addition, the efficiency of this
torque component compared to the gas torque depends on the dust-to-gas
mass ratio/distribution. We summarize our results in a first torque
map for dusty disks which we present in Fig.\,\ref{fig:fig2}.

\section{Numerical simulations}
\label{sec:numerical}

We perform numerical simulations using the
FARGO3D\footnote{\href{http://fargo.in2p3.fr}{http://fargo.in2p3.fr}}
code \citep{Benitez-Llambay2016}, which solves the equations for fluid
dynamics using finite-difference upwind, dimensionally split methods,
combined with the FARGO algorithm \citep{Masset2000} for the orbital
advection and a fifth order Runge-Kutta integrator for planetary
orbits. We extended the code to deal with multiple fluids, and treat
the dust particles as a pressureless fluid (Ben\'itez-Llambay et al.,
in prep).

The basic equations of our model are those given in
\cite{Weber2018}. We employ a disk model that corresponds to the
steady-state solution of an $\alpha-$disk \citep{Shakura1973}. The
background density and temperature are power laws in radius with
indices $-1/2$ and $-1$, respectively.
The aspect ratio of the disk is set to $h = 0.05$ and the
$\alpha$-parameter is set to $\alpha=3\times 10^{-3}$.
For the dust component we adopt the steady-state solution described in
\cite{Takeuchi2002}.
Because dust feedback is not considered and we parametrize the drag
force via the Stokes number, the absolute value for the gas and dust
surface densities, $\Sigma_{\rm g}$ and $\Sigma_{\rm d}$,
respectively, are not relevant. For the sake of definiteness, we set
both to unity. We use a locally isothermal equation of state for the
gas.

\begin{figure}
  \includegraphics[width=\columnwidth]{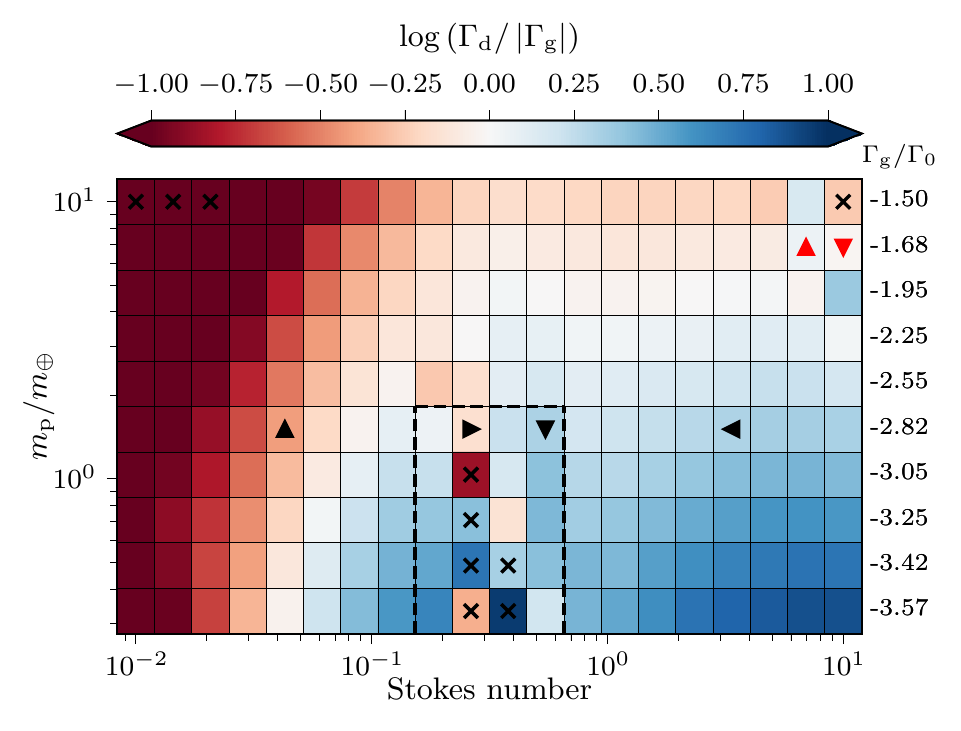}
  \caption{Torque map summarizing the numerical exploration. Positive
    dust torques are found in the majority of the cases, except those
    marked with crosses, for which we use the absolute value of the
    torque. There are two regions where dust torques exceed gas
    torques, leading to outward migration. For small Stokes numbers
    (gas-dominated regime) the dust torque is produced by small-scale
    asymmetries close to the planet. For large Stokes numbers
    (gravity-dominated regime), the dust torque is due to the dust
    hole behind the planet. The dust torques were obtained considering
    a $1\%$ dust-to-gas mass ratio. Numbers to the right provide the
    gas torque normalized by $\Gamma_0 = (m_{\rm
      p}/M_\star)^2{\Sigma_\mathrm{g}}_0 r_{\rm p}^4 \Omega_{\rm
      p}^2/h^{2}$.  Because the dust torque scales with the dust
    surface density, for a given dust-mass distribution, the net
    torque is obtained by direct multiplication by the dust-to-gas
    ratio per bin.  The black (red) triangles correspond to the
    simulations presented in Fig.\ref{fig:fig1}
    (Fig.\ref{fig:fig3}). The rectangle delimited by the dashed line
    corresponds to high-resolution simulations in the transition
    region between the gas and gravity-dominated regimes.
    \label{fig:fig2}}
\end{figure}

We implement a non-uniform 2D polar grid radially refined close to the
planet according to the mesh density function
$\psi(r) = 1/r + \xi/[\left(r-r_\mathrm{p}\right)^2+\xi^2]$, 
where the parameter $\xi$, taken to be 0.1, sets the relative density
of cells close to the planet location $r_{\rm p}$, which we consider
fixed in a frame rotating with angular frequency $\Omega_{\rm p}$. The
mesh has 768 radial zones in the domain $r \in [0.48, 2.08]\,r_{\rm
  p}$, which results in a cell size of $5\times 10^{-4} r_{\rm p}$ at
the planet position, allowing us to resolve both the horseshoe orbits
of the smallest planet considered, $m_{\rm p}=10^{-6} M_\star$ (with
$M_\star$ the stellar mass), and the pressure scale for all the
Lindblad resonances far away from the planet. The mesh is azimuthally
divided in 3072 uniform sectors. In the cases limited by the dashed
line in Fig.\,\ref{fig:fig2}, we increased the resolution by using
12288 azimuthal zones and setting $\xi = 0.05$, which results in a
resolution of $\sim 2.7\times10^{-4} r_{\rm p}/\mathrm{cell}$ at the
planet location.

We adopt damping buffer zones \citep{Val-Borro2006} for the radial
velocity, in rings extending from the inner (outer) boundaries to the
radius where the angular frequency of the disk matches 2/3 (3/2) of
the angular frequency at the boundary. The damping timescale is chosen
equal to $0.3$ local orbital periods \citep{Benitez-Llambay2016a}.
The smoothing length for the planet potential is equal to its Hill
radius.  We explore the problem by setting a two dimensional,
logarithmic grid in the parameter space spanned by the planetary mass
and the Stokes number, defining 10 bins in the range $m_{\rm p} \in
[10^{-6},3\times10^{-5}]\,M_\star$ and 20 in the Stokes number range
$[10^{-2},10]$.
The total integration time is equal to 50 planetary orbits. The
majority of the simulations reach steady state long before this time
span. A systematic exploration involving hundreds or thousands of
orbits at lower resolutions confirms that longer integration times do
not alter the torque measurements that we report below.

\section{Torque map for dusty disks}

In order to assess the importance of the dust torque compared to the
gaseous one, we summarize our measurements in the torque map shown in
Fig.\,\ref{fig:fig2}. We plot the relative strength of the dust
torque, $\Gamma_{\rm d}$, to the gas torque, $\Gamma_{\rm g}$, for all
of the simulations carried out. The torques were obtained as
instantaneous measurements at 50 orbits. This is justified because the
values of the torques vary little with respect to their mean.

Valuable insight for understanding the physical processes determining
the global morphology of the torque map can be obtained by analyzing
the dust dynamics in the vicinity of the planet. With this goal, we
show in Fig.\,\ref{fig:fig1} the dust surface density perturbation and
torque densities for different Stokes number at constant planetary
mass. We overlay the dust streamlines, showing the expected horseshoe
orbits altered by a strong radial velocity induced by the drag force
\citep[cf.][]{Ogilvie2006d, Morbidelli2012}.
Particles approaching the planet experience a strong torque, which
increases their semi-major axis and eccentricity, a process well
captured by the impulse approximation \citep{Lin1986}.

As the Stokes number increases (from left to right),
Fig.\ref{fig:fig1} displays a clear trend. The drag force decreases
and, as expected, the particles experiment larger variations in their
orbital elements.
In the limit of perfect coupling (Stokes number $\to 0$), the dust is
a perfect tracer of the gas, while in the regime where the dust is
dynamically decoupled (Stokes number $\to \infty$), the particles
trace the test particle orbits in the restricted three-body
problem. This can clearly be seen by comparing the size of the
horseshoe orbits in panels (a) and (d)
\citep[see][]{Paardekooper2009b}.

\newpage 

\subsection{Gas and Gravity-dominated Regimes}

We identify two different regimes leading to positive dust torques:
the \textit{gas}-dominated and \textit{gravity}-dominated regimes,
which are connected by a transition regime, in which the dust torque
may decrease in magnitude and, for some cases, reverse sign.

\begin{figure}
  \includegraphics[width=\columnwidth]{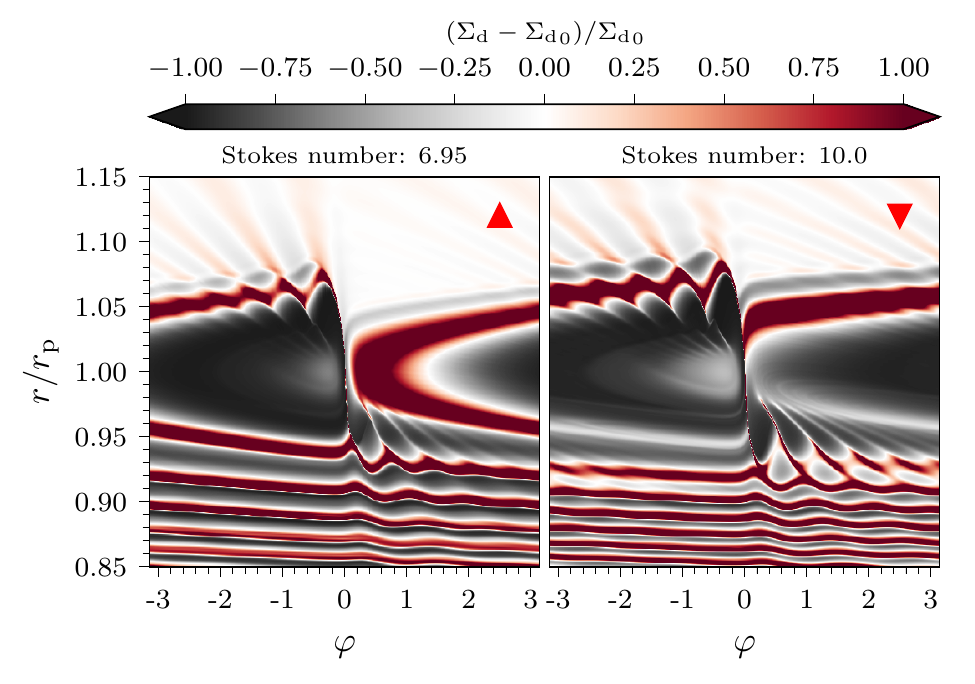}
  \caption{Relative dust surface density for two different Stokes
    numbers, corresponding to the red triangles in
    Fig.\,\ref{fig:fig2}. Left panel: an overdense stream is able to
    enhance the positive torque when approaching the planet from its
    front. Right panel: For larger Stokes numbers, this stream is
    scattered away by the planet, transforming the hole into a gap
    and, in this case, decreasing the dust-torque.}
  \label{fig:fig3}
\end{figure}

\emph{Gas-dominated regime (a $\&$ e) \/--} Dust scattering is only
efficient very close to the planet, accumulating an incident tube of
particle streamlines along a ring-shaped overdense stream.
In this process the eccentricity of the scattered dust is quickly
damped. This overdensity corresponds to the limit of a protected
region, which, for the sake of simplicity, we call the \textit{dust
  hole}.
The dust hole is isolated from the incoming beam of particles and,
because it is draining out its dust content, it becomes
underdense. Because an underdense counterpart does not exist in front
of the planet, this enhances the possibility that a positive torque
develops.
When inspecting the torques close to the planet, a torque dipole is
observed as well as two small filaments, the origin of which can be
tracked down to a stagnation point located behind the planet.
Dust concentrates in a filament that crosses the Hill radius of the
planet. Interestingly, this asymmetry produces a positive torque, a
robust result to variations in the planet mass and the Stokes number.
We find the dust torques arising in this region of parameter space to
be positive and gradually increasing with Stokes number. In addition,
its ratio to the gaseous torque decreases with planetary mass.
As the Stokes number increases further, the dust hole shrinks and, for
very low-mass planets, it may eventually close, leading to the
transition regime.
We note that in the small Stokes number regime, it may be possible to
gain insight into the dynamical processes determining the dust torque
by employing the formalism introduced in \citet{Lin2017}.

\emph{Transition regime (b $\&$ f) \/--} As the Stokes number
increases (resulting in weaker drag), the hole develops and expands.
The interplay between the scattered dust flow induced by the planet
and the drag force produced by the aerodynamic friction with the gas
effectively generates a barrier.  Particles arriving from the ambient
flow cannot enter while dissipative forces expel those inside. This is
illustrated by the white streamline in panel (d) of
Fig.\,\ref{fig:fig1}.
While the drift speed increases, the drag force remains large enough
so as to prevent efficient scattering to occur and the dust tends to
accumulate behind and closer to the planet.
The dust overdensity behind the planet now competes against the
overdense filament in front of the planet, effectively reducing the
net dust torque.
As the Stokes number increases further, the overdense lobe behind the
planet rapidly expands leading to an abrupt, but continuous,
transition into the gravity-dominated regime.
Depending on the mass of the planet, this can reduce the positive
torque and even change its sign.
When inspecting the torque distribution, we observe that small planets
seem to be unable to open a hole in order to reduce the excess of
negative torque, leading to a net negative dust torque. In general,
the less massive the planet, the more negative the torque can become.
We find that, depending on the mass of the planet, the value of the
torque in this regime can be very sensitive to the dust distribution
close to the planet and the shape of the hole.  As the hole seems to
be absent, or is very tiny for small planets, very high resolution is
imperative in order to correctly resolve this regime. For larger
Stokes numbers, the hole continues expanding both radially and
azimuthally, and the dust torque enters the gravity-dominated regime.
 
\begin{figure*}
  \centering
  \includegraphics[width=0.9\textwidth]{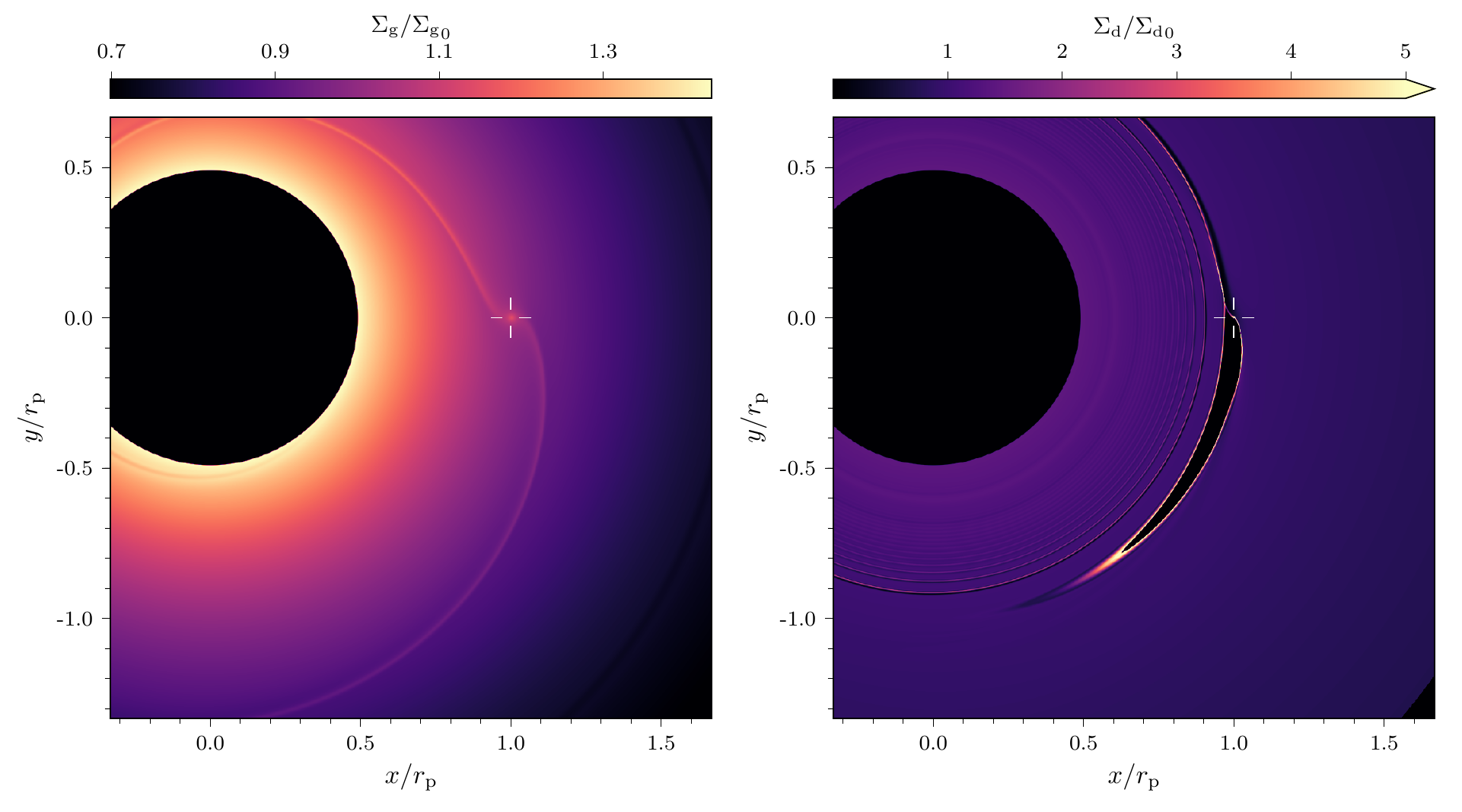}
  \caption{Cartesian representation of the gas and dust surface
    density corresponding to the simulation presented in panel (d) in
    Fig.\,\ref{fig:fig1}. The white crosshair marks the planet's
    position.}
  \label{fig:fig4}
\end{figure*}

\emph{Gravity dominated regime (c $\&$ g)-(d $\&$ h) \/--} This regime
is characterized by significant changes in the dust dynamics, in which
the semi-width of the horseshoe orbits is modified from $\sim r_{\rm
  p} \sqrt{m_{\rm p}/(hM_{\star})}$ to $\sim 2.5 r_{\rm p}
\sqrt[3]{m_{\rm p}/(3M_\star)}$ \citep{Paardekooper2009b}, as well as
a change in the dust drift speed.
For larger Stokes numbers (d), depending on the mass of the planet,
epicyclic motion starts to be evident.  Naturally, because there
exists a hole behind the planet and an overdense stream ahead of it,
the net torque becomes positive. The larger the Stokes number is, the
larger the azimuthal size of the hole becomes, providing a natural
scaling for the torque with the Stokes number.
The general picture, however, is not simple. For example, while the
size of the hole increases, because the disk is azimuthally periodic
the overdense stream approaches the planet from the front, thus
enhancing the torque. This feature is observed for intermediate masses
and large Stokes numbers, as shown in the left panel of
Fig.\,\ref{fig:fig3}.
For large planet masses and Stokes numbers, the planet perturbations
are such that a dust particle is not able to drift before it gets
scattered again by the planet, creating a dust barrier that leads to
the formation of a gap in the dust component.
This kind of configuration is shown in the right panel of
Fig.\,\ref{fig:fig3}.
It is worth noticing that the physical processes responsible for
producing the dust torque in this regime are similar to those driving
type-III migration \citep[e.g.][]{Masset2003, Peplinski2008a}. The
underdense dust hole is analogous to the \textit{gas mass deficit} in
the co-orbital region in type-III migration. In both of these cases,
the size and shape of the mass deficit depends on the planet mass and
the drift speed. We thus anticipate that the concepts developed in the
framework of type-III migration could be adapted to investigate the
dynamical processes shaping the dust hole.  We illustrate the gas and
dust surface densities in Cartesian coordinates for one of our
simulations in Fig.\,\ref{fig:fig4}.

\smallskip

It is worth noticing that there exist similarities between our
findings and those presented by \cite{Capobianco2011a}, who
investigated the potential for planetesimals to alter the migration
history of small mass cores embedded in a gaseous disk. In order to
investigate this so-called planetesimal-driven migration, their study,
based on the N-body approach, focuses on the (cumulative effects of
the) torque exerted by close encounters subject to a simple
aerodynamic model. This is related, but different, to our
approach. Here, we consider the torque exerted by the whole dust
distribution in steady state taking into account the gas
dynamics. This makes it possible for us to assess the net effect in
terms of the resulting global asymmetry in the dust surface
density. Nevertheless, \cite{Capobianco2011a} also found that
planetesimal-driven outward migration is possible for small mass
planets. In spite of the different scales and approaches involved in
studying these two problems, this seems to be an indication that the
mechanisms involved share important similarities. This calls for
further investigation in order to establish any potential connections
on solid grounds.

\smallskip

\subsection{Assessment of the Robustness of our Findings}

The precise value obtained when measuring the dust torque is sensitive
to the smoothing length.
As the smoothing necessary to produce results comparable to a 3D
simulation (which is probably sensitive to the dust distribution in
the vertical direction and the Stokes number) is not known, we adopted
a smoothing length that scales with the quantity that seems to be
dynamically relevant for dust, i.e., the Hill radius. We note,
however, that the effective smoothing in the gas-dominated regime may
be larger. This is because in the well-coupled regime, the smoothing
length should tend toward that of the gas component ($\epsilon/r_{\rm
  p}=0.6 h$, \cite{Mueller2012}).
We have checked that measuring the torque with different smoothing
lengths does not alter significantly the trends presented in
Fig.\,\ref{fig:fig2}.
In particular, the dust torque in the gravity-dominated regime is very
robust, because it arises from a large-scale asymmetry. This is,
however, not necessarily the case in the gas-dominated regime, where
the dust torque results from a more subtle asymmetry close to the
planet. In order to minimize spurious measurements due to unresolved
dynamics, we have conservatively cut off the inner half of the Hill
sphere when computing torques.
A more thorough assessment of the robustness of our results will
require 3D high-resolution simulations, where the smoothing length is
not a free parameter.

\smallskip

The surface density profile has an impact on the gaseous corotation
torque \citep[e.g.,][]{Tanaka2002a}, which is usually a positive
quantity and vanishes for surface density profiles $\propto
r^{-3/2}$. This contribution makes the total gas torque less negative,
thus enabling the dust torque to slow down or even reverting the
migration \citep{Paardekooper2010}. We have used $\Sigma_{\rm g}
\propto r^{-1}$, which leads to a rather conservative corotation
torque. We anticipate that the effect we describe is prone to be
enhanced in more realistic disks, where the corotation torque may
increase considerably.

\smallskip

We have modeled the dust as a pressureless fluid. This approach may
break down when the dynamics is governed by crossing orbits in
low-density regions. In this regime, further studies employing
Lagrangian particles are necessary to ultimately validate our
findings. It is worth mentioning, however, that the streamlines
characterizing the dust dynamics in our models share many key
properties with those presented using particles
\citep[e.g.,][]{Morbidelli2012}.

\section{Future Work and Implications}

In this Letter, we have provided a first assessment of the potential
role of dust dynamics on the migration history of low mass planetary
embryos. We have identified asymmetries in the dust-density
distribution in the vicinity of the planet orbit that may play an
important role in determining the net torque on the planet
(Fig.\,\ref{fig:fig1}). We have carried out systematic measurements of
these torques and summarized the outcome of our study in a torque map
(Fig.\,\ref{fig:fig2}).  This will be useful for incorporating the
effect of dust dynamics on migration into planet population synthesis
models. We briefly discuss below avenues for future developments and
some of the potential implications of our findings.

The efficiency of the dust torque is determined by the relative flow
between the dust and the planetary embryo.  There are several
processes, which we have not considered in this first work, that may
play a significant role in more refined models of the early evolution
of planetary cores. In particular, future models should account for
dust feedback, accretion onto the planetary embryo, planetary
migration, and the influence of other planets.
We also note that the dust torque in the gas-dominated regime is
likely to be sensitive to the gaseous equation of state, because of
its impact on the gas flow in the vicinity of the planet. Further
studies are necessary in order quantify and asses the robustness of
dust torques in this regime under such modifications.

The strong asymmetries in the dust distribution will alter the opacity
of the gas and thus its thermal structure close to the planet. This
may have an impact on the thermal torques \citep[e.g][]{Lega2014,
  Benitez-Llambay2015a, Masset2017b}. Building a unified picture of
the migration history of low-mass planets will require modeling dust
dynamics, accretion, and thermodynamics self-consistently
\citep[e.g.,][]{Chrenko2017}.

The dust torque scales with the dust-mass content, which in addition
scales with the metallicity of the disk. Therefore, we predict that
outward migration of low-mass embryos is likely to be more effective
in systems with higher-metallicity, enhancing the occurrence of
distant giant planet in higher metallicity systems.

In order to fully assess the role of dust torques on planet migration
it is necessary to consider realistic dust-size/mass
distributions. Nevertheless, the dust torque seems a good candidate
for slowing down or even revert the migration of low-mass planetary
cores.  We thus anticipate that dust-driven migration could play a
dominant role during the formation history of planets, including those
in our own Solar System.

\acknowledgements

We thank the anonymous referee for a timely and constructive report,
which helped us improve the manuscript.
We thank Leonardo Krapp, Alejandro Benitez-Llambay \& Ximena Ramos for
useful discussions and Frederic Masset, Tobias Heinemann and Oliver
Gressel for useful comments.
We thank Colin McNally for his valuable input based on a preprint
version of this Letter.
This project has received funding from the European Union's Horizon
2020 research and innovation programme under grant agreement No.
748544.
The research leading to these results has received funding from the
European Research Council (ERC) under the European Union's Seventh
Framework programme (FP/2007-2013) under ERC grant agreement
No. 306614 and under the European Union's Horizon 2020 research and
innovation programme (grant agreement No 638596).
This work was supported by a research grant (VKR023406) from VILLUM
FONDEN.
We acknowledge PRACE for awarding us access to MareNostrum at
Barcelona Supercomputing Center (BSC), Spain.
Computations were performed at the HPC center of the University of
Copenhagen.

\software{IPython \citep{Perez2007}, NumPy \citep{Walt2011},
  Matplotlib \citep{Hunter2007}}

\end{document}